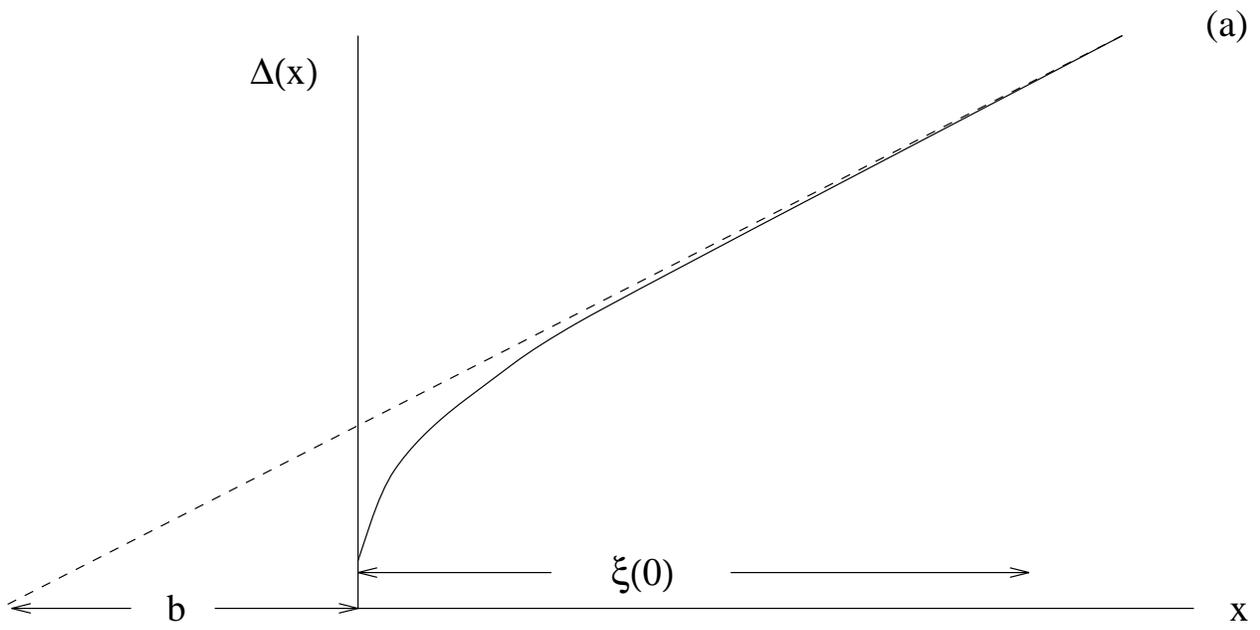

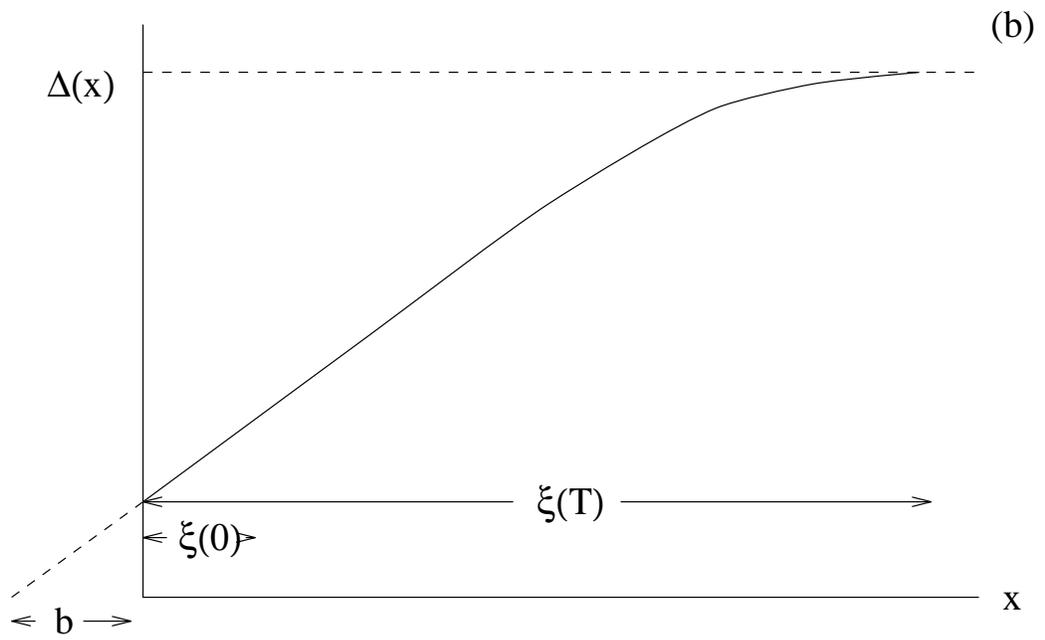

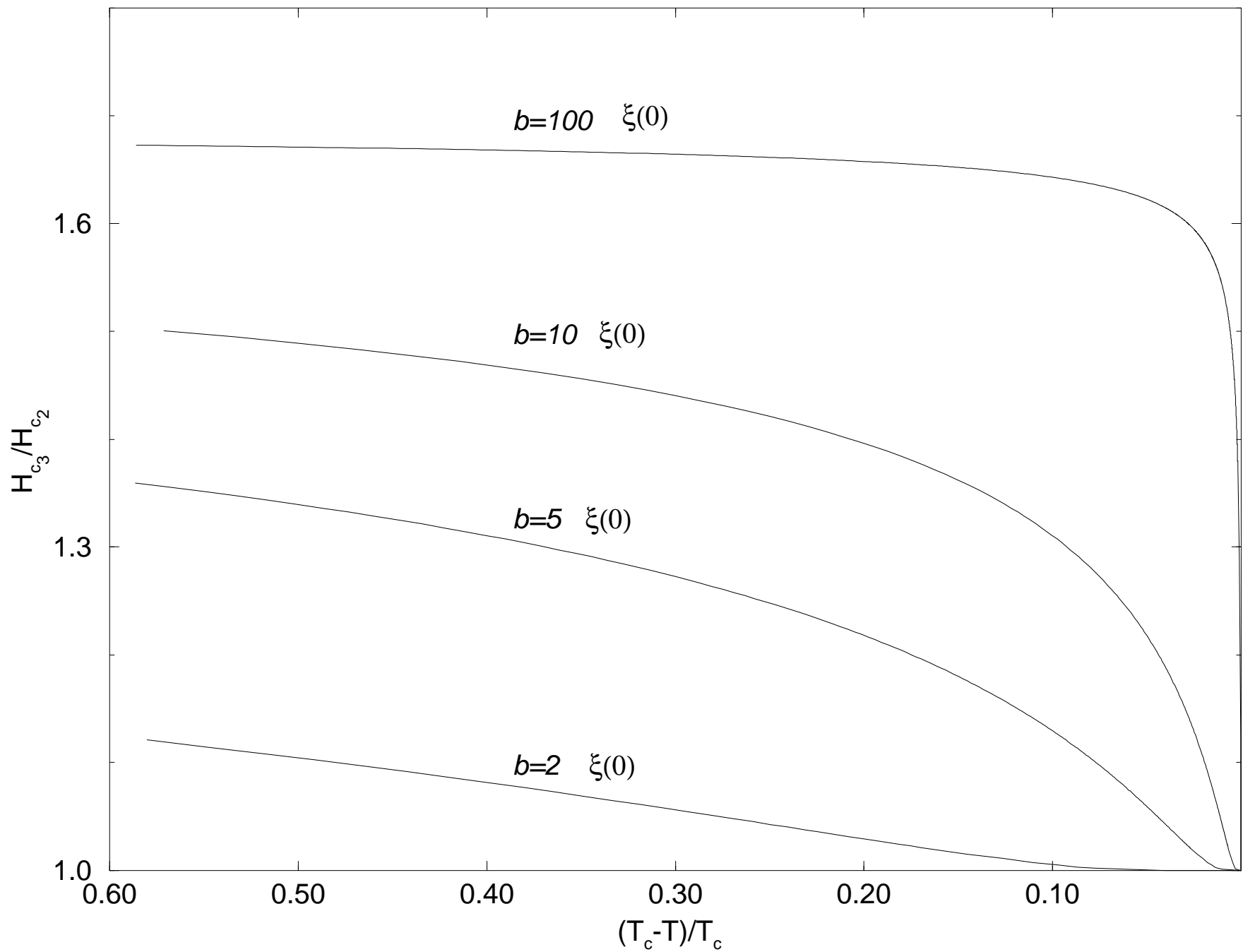

# Effect of diffusive boundaries on surface superconductivity in unconventional superconductors

D.F. Agterberg and M.B. Walker

*Department of Physics, University of Toronto, Toronto, Ontario, Canada, M5S 1A7*

(May 7, 1996)

## Abstract

Boundary conditions for a superconducting order parameter at a diffusive scattering boundary are derived from microscopic theory. The results indicate that for all but isotropic gap functions the diffusive boundary almost completely suppresses surface superconductivity in the Ginzburg-Landau regime. This indicates that in anisotropic superconductors surface superconductivity can only be observed for surface normals along high symmetry directions where atomically clean surfaces can be cleaved.

74.60.Ec, 74.70.Tx

Typeset using REVTEX

Superconductivity in UPt$_3$ is believed to be described by an unconventional order parameter which has two complex components $\eta_1$ and $\eta_2$ [1–5]. This description allows for an explanation of the many unique experimental features associated with superconductivity in this hexagonal material. For example, the upper critical field for fields in the basal plane ($H_{c_2}^{ab}$) as a function of temperature displays a kink at a temperature $T^*$ [6]. At temperatures below $T^*$ one component of the order parameter orders at $H_{c_2}^{ab}$, while for temperatures above $T^*$ the other component orders at $H_{c_2}^{ab}$. The observation and experimental investigation [6–8] of surface superconductivity in UPt$_3$ has lead to an examination of this phenomenon in unconventional superconductors [9–11]. In contrast to $H_{c_2}^{ab}$, the upper critical field for surface superconductivity for fields in the basal plane ($H_{c_3}^{ab}$) exhibits no anomaly with temperature [8]. This can be understood if one component of the order parameter is suppressed at the surface, allowing surface superconductivity to occur only with the other component [9,11]. It is therefore important to understand under what conditions the superconducting order parameter is suppressed at a boundary. Microscopic calculations for boundary conditions at specular reflecting surfaces have been conducted [10,12]. It is also interesting to examine the effects of diffusive boundaries on surface superconductivity. It is known that rough surfaces are pair breaking for anisotropic superconductors [13,14] and therefore are expected to suppress surface superconductivity. Here we investigate how strong such a suppression will be for general order parameter symmetries. First, we examine the solution of the isotropic Ginzburg-Landau model with general boundary conditions to gain an understanding of the effects of the boundary conditions on surface superconductivity. Then we use a weak coupling microscopic theory to calculate the boundary conditions at a diffusive scattering boundary for general order parameter symmetries.

The Ginzburg-Landau free energy density for an isotropic superconductor in an external magnetic field **H** is

$$F = \alpha_0(T - T_c)|\psi|^2 + \kappa(\mathbf{D}\psi) \cdot (\mathbf{D}\psi)^* + \mathbf{h}^2/8\pi - \mathbf{h} \cdot \mathbf{H}/4\pi \tag{1}$$

where $\mathbf{D} = \boldsymbol{\partial} - (i2e/\hbar c)\mathbf{A}$, and $\mathbf{h} = \boldsymbol{\partial} \times \mathbf{A}$. We have only kept terms to order $\psi^2$ since we are interested in determining the upper critical field. In the presence of a surface the following surface free energy density is added

$$F_S = g|\psi|^2. \tag{2}$$

The Ginzburg-Landau equations are

$$\alpha\psi = \kappa \mathbf{D} \cdot \mathbf{D}\psi \qquad \mathbf{h} = \mathbf{H} \tag{3}$$

where $\alpha = \alpha_0(T - T_c)$ with the boundary condition $\mathbf{n} \cdot \mathbf{D}\psi|_{surface} = (1/b)\psi|_{surface}$ where $\mathbf{n}$ is the surface normal and $b = \kappa/g$ is the extrapolation length and is described in Fig. 1 (also see Ref. [18]). For an applied magnetic field orthogonal the the surface normal this equation can be solved by following the method of Saint James [15]. The solution is

$$\frac{H_{c_3}}{H_{c_2}} = \frac{1}{\mu^2(l) - l^2} \tag{4}$$

where $H_{c_2} = 2\pi\phi_0/\xi^2(T)$, $\phi_0$ is the elementary flux quantum, $\xi(T) = \sqrt{-\kappa/\alpha}$ is the coherence length, and $\mu(l)$ is given by

$$\int_0^\infty (2u - \mu - l)e^{-(u-\mu)^2} u^{-(1+\mu^2-l^2)/2} du = 0. \tag{5}$$

where $l = (H_{c_2}/H_{c_3})^{1/2}\xi(T)/b$. The ratio $H_{c_3}/H_{c_2}$ as a function of temperature is given in Fig. 2.

Microscopic calculations are required to determine the ratio $b/\xi(0)$. Such calculations have been performed for general order parameter symmetries in the presence of a specular reflecting surface in a weak coupling model with a spherical Fermi surface [10,12]. Here similar calculations are performed for a diffusive scattering surface (this has previously been done for isotropic [16], weakly anisotropic [17], and $p$-wave [13] order parameters). The correlation function method developed by deGennes [18,13] and extended to unconventional superconductors by Sigrist and Ueda [19] is used. We consider a weak coupling model with a spherical Fermi surface and assume that there is no spin flip scattering at the surface. The development of the formalism initially parallels that of Sigrist and Ueda [19].

The Hamiltonian is

$$H = \sum_{\substack{k,k' \\ s,s'}} \langle ks|H_0|k's'\rangle c^\dagger_{ks} c_{k's'} + \sum_{\substack{k,k',q \\ s_1,s_2,s_3,s_4}} V_{s_1,s_2,s_3,s_4}(\mathbf{k},\mathbf{k}') c^\dagger_{\mathbf{q}/2-\mathbf{k},s_1} c^\dagger_{\mathbf{q}/2+\mathbf{k},s_2} c_{\mathbf{q}/2+\mathbf{k}',s_3} c_{\mathbf{q}/2-\mathbf{k}',s_4} \quad (6)$$

where $H_0$ is the single particle Hamiltonian including the interaction due to the boundary and $c_{\mathbf{k},s}$ destroys a free electron with Momentum $\mathbf{k}$ and spin $s$. Defining $G_{s,s'}(\mathbf{k},\mathbf{k}',\tau) = -\langle T_\tau\{c_{\mathbf{k},s}(\tau), c^\dagger_{\mathbf{k}',s'}(0)\}\rangle$ and $F^\dagger_{s,s'}(\mathbf{k},\mathbf{k}',\tau) = \langle T_\tau\{c^\dagger_{\mathbf{k}',s'},(\tau)c^\dagger_{\mathbf{k},s}(0)\}\rangle$ where $c_{\mathbf{k},s}(\tau) = \exp^{H\tau} c_{\mathbf{k},s}\exp^{-H\tau}$ and $T_\tau$ is the imaginary time ordering operator and finding the equations of motion for these averages leads to the linearized gap equation

$$\Delta_{s_1,s_2}(\mathbf{q},\mathbf{k}) = -k_B T \sum_{\omega_n} \sum_{\substack{\mathbf{k}',\mathbf{k}'',\mathbf{q}' \\ s_3,s_4,s_5,s_6}} V_{s_2,s_1,s_3,s_4}(\mathbf{k},\mathbf{k}') G^0_{s_3,s_5}\left(\mathbf{k}'+\mathbf{q}/2, \mathbf{k}''+\mathbf{q}'/2; i\omega_n\right)$$
$$\times G^0_{s_4,s_6}\left(-\mathbf{k}'+\mathbf{q}/2, -\mathbf{k}''+\mathbf{q}'/2; -i\omega_n\right) \Delta_{s_5,s_6}(\mathbf{k}'',\mathbf{q}') \quad (7)$$

where $\omega_n = \pi k_B T(2n+1)$ are the Matsubara frequencies and the normal state electron Greens function $G^0_{s,s'}(\mathbf{k},\mathbf{k}',i\omega_n)$ is given by the Fourier transform of

$$G^0_{s,s'}(\mathbf{R},\mathbf{R}',i\omega_n) = \sum_\nu \frac{\phi^*_{\nu,s}(\mathbf{R})\phi_{\nu,s'}(\mathbf{R}')}{i\omega_n - \epsilon_\nu} \delta_{s,s'} \quad (8)$$

where $\phi_{\nu,s}(\mathbf{R})$ are the eigenfunctions of the single particle Hamiltonian $H_0$. We assume that the interaction can be written in the weak coupling form

$$V_{s_1,s_2,s_3,s_4}(\mathbf{k},\mathbf{k}') = \sum_{\Gamma,m} g(\Gamma) \Delta_{s_2,s_1}(\Gamma,m,\mathbf{k_F}) \Delta^\dagger_{s_3,s_4}(\Gamma,m,\mathbf{k'_F}) \quad (9)$$

where $\epsilon(\mathbf{k})$ and $\epsilon(\mathbf{k}')$ are restricted to lie within an energy $\epsilon_c$ of $\epsilon_F$, $\Gamma$ refers to an irreducible representation of the point group and $m$ to the basis of the $\Gamma$ representation. If only one representation is important then the gap matrix can be written as $\hat{\Delta}(\mathbf{R},\mathbf{k_F}) = \sum_m \eta_m(\mathbf{R})\hat{\Delta}_m(\mathbf{k_F})$. Fourier transforming the linearized gap equation with respect to the center of mass variables $\mathbf{q}$ and $\mathbf{q}'$, substituting the above forms for the gap function and the potential, and using the following orthogonality condition for the gap matrix

$$\frac{1}{4\pi k_F^2} \int_S d^2k \, Tr[\hat{\Delta}_l(\mathbf{k_F})\hat{\Delta}_m(\mathbf{k_F})] = 2\delta_{m,l} \quad (10)$$

where the integral is over the Fermi surface, gives the following equation for the order parameter $\eta_i(\mathbf{R})$

$$\eta_i(\mathbf{R}) = \int d^3 R' K_{ij}(\mathbf{R}, \mathbf{R}')\eta_j(\mathbf{R}') \tag{11}$$

with $K_{ij}(\mathbf{R}, \mathbf{R}')$ given by [19]

$$K_{ij}(\mathbf{R}, \mathbf{R}') = \frac{gk_BT}{2} \sum_{\omega_n} \int_0^\infty dt \int d\epsilon i \frac{\exp^{-(|\omega_n|-i\epsilon)t}}{i\omega_n - \epsilon} \tag{12}$$

$$\sum_\nu \left\langle \nu \left| tr \hat{\Delta}_i^\dagger \left[\frac{m}{2k_F}\mathbf{J}(\mathbf{R})\right] \hat{\Delta}_j \left[\frac{m}{2k_F}\mathbf{J}(\mathbf{R}', t)\right] \right| \nu \right\rangle \delta(\epsilon - \epsilon_\nu) \tag{13}$$

where $\mathbf{J}(\mathbf{R})$ is the current operator. We use the semiclassical and weak coupling approximations, which entail $\sum_\nu \langle tr\hat{\Delta}_i^\dagger \hat{\Delta}_j \rangle \delta(\epsilon_v - \epsilon) \approx N(0)\langle tr\hat{\Delta}_i^\dagger \hat{\Delta}_j \rangle_{\epsilon=\epsilon_F,classical}$ [18,13] where $N(0)$ is the density of states at the Fermi surface, to arrive the following from for the kernel [19]

$$K_{ij} = gN(0)\pi k_B T \sum_{\omega_n} \int_0^\infty dt \exp^{-2|\omega_n|t} \left\langle tr\hat{\Delta}_i^\dagger \left[\frac{m}{2k_F}\mathbf{J}(\mathbf{R})\right]\hat{\Delta}_j\left[\frac{m}{2k_F}\mathbf{J}(\mathbf{R}',t)\right] \right\rangle_{\epsilon_F,classical} \tag{14}$$

where the expectation value is an average in a canonical ensemble for an electron with momentum on the Fermi surface. In the presence of a single boundary, the kernel has two parts, a direct contribution ($K^d$) which is the contribution when no boundary is present and a contribution due solely to the scattering at the boundary ($K^r$). For simplicity, we assume a spherical Fermi surface in which case the direct contribution is given by Sigrist and Ueda to be [19]

$$K_{ij}^d(\mathbf{R}) = \frac{gN(0)k_BT}{2v_F} \sum_{\omega_n} \frac{tr\left[\hat{\Delta}_i^\dagger[\frac{\mathbf{R}}{R}]\hat{\Delta}_j[\frac{\mathbf{R}}{R}]\right]}{R^2} \exp^{-\frac{2|\omega_n|R}{v_F}}. \tag{15}$$

The transition temperature is given by the condition $\int d^3 R K_{ii}^d(\mathbf{R}) = 1$ where $i$ corresponds to only one component of the order parameter (for a more detailed discussion of this point see [18]). For a diffusive boundary the expectation value in $K_{ij}^r$ is given by

$$\left\langle tr\hat{\Delta}_i^\dagger \hat{\Delta}_j \right\rangle = (4N(0)k_F^2)^{-1} \int_{p_z<0} \frac{d^3p}{(2\pi)^3} \int_{p_z'>0} d^3p' \frac{p_z'}{\pi p'} \delta^3[\mathbf{R}' - \mathbf{R}^\perp + \mathbf{p}^\perp \frac{R_z}{p_z} - \mathbf{p}'(\frac{t}{m} + \frac{R_z}{p_z})]$$

$$\delta(p - p')\delta(\frac{p^2}{2m} - \epsilon_F)tr\left[\hat{\Delta}_i^\dagger(\mathbf{p})\hat{\Delta}_j(\mathbf{p}')\right]. \tag{16}$$

In this equation a quasiparticle that had initial position $\mathbf{R} = (\mathbf{R}_\perp, R_z)$ ($z$ is the component along the surface normal) and initial momentum $\mathbf{p} = (\mathbf{p}_\perp, p_z)$ has been scattered by the surface emerging with momentum $\mathbf{p}'$. The $p'_z/\pi p'$ represents the probability of emerging from the surface with momentum $\mathbf{p}'$ (note this is independent of $\mathbf{p}$) and the 3D delta function gives the time dependent position $\mathbf{R}'(t)$ of the quasiparticle given that at time $t = -mR_z/p_z$ the quasiparticle is at the surface with position $\mathbf{R}_\perp - \mathbf{p}_\perp R_z/p_z$ (these correspond to the position on the surface and the time required to reach the surface for a quasiparticle with initial position and momentum given by $\mathbf{R}$ and $\mathbf{p}$)(see Ref. [16] for a discussion of diffusive scattering for the case of isotropic superconductors). We assume that the order parameter varies only in the direction along the surface normal and therefore wish to determine $\int d^2(R_\perp - R'_\perp) K_{ij}(\mathbf{R}, \mathbf{R}') = K_{ij}(z, z')$. The resulting kernel is

$$K_{ij}(z, z') = \frac{gN(0)k_b T \pi}{2 v_F} \sum_n \left\{ \int_0^1 \frac{ds}{s} \exp^{-2\frac{|\omega_n||z-z'|}{v_F s}} F_{ij}(s) \right.$$
$$\left. + \pi^{-1} tr \left[ \int_0^1 ds \exp^{-2\frac{|\omega_n||z|}{v_F s}} \hat{\Delta}_i^\dagger(s) \right] \times \left[ \int_0^1 ds \exp^{-2\frac{|\omega_n||z'|}{v_F s}} \hat{\Delta}_j(-s) \right] \right\} \quad (17)$$

where $F_{ij}(s) = (1/2) tr \int_0^{2\pi} d\phi [\hat{\Delta}_i^\dagger(s,\phi)\hat{\Delta}_j(s,\phi) + \hat{\Delta}_i^\dagger(-s,\phi)\hat{\Delta}_j(-s,\phi)]$, $\hat{\Delta}_i(s) = \int_0^{2\pi} d\phi \hat{\Delta}_i(s,\phi)$, and $\hat{\Delta}_i(\phi, s)$ is given by setting $\mathbf{k} = (\sqrt{1-s^2}\cos\phi, \sqrt{1-s^2}\sin\phi, s)$ in $\hat{\Delta}(\mathbf{k})$. To obtain Eq. 17 the following was used

$$\int_0^{2\pi} d\phi \hat{\Delta}_i^\dagger(\phi, \rho, z) \hat{\Delta}_j(\phi, \rho, z) = \int_0^{2\pi} d\phi \hat{\Delta}_i^\dagger(\phi, \rho, -z) \hat{\Delta}_j(\phi, \rho, -z) \quad (18)$$

where $\hat{\Delta}(\phi, \rho, z)$ is given by setting $\mathbf{k} = (\rho\cos\phi, \rho\sin\phi, z)$ in $\hat{\Delta}(\mathbf{k})$. Eq. 18 arises because both $\hat{\Delta}_i$ and $\hat{\Delta}_j$ transform identically under parity. A similar development for a specularly reflecting surface gives the same result as Samokhin [10].

To proceed we consider the $i = j$ contributions only (these are frequently the only contributions along high symmetry directions). After introducing $x = z/\xi_0$, where $\xi_0 = \hbar v_F / 2\pi k_b T_c$, the integral equation for the order parameter becomes

$$\eta_i(x) = \int_0^\infty dx' \tilde{K}(x, x') \eta_i(x') \quad (19)$$

where $\tilde{K} = (\hbar v_F / 2\pi k_b T_c) K_{ii}$. It can be verified that $\eta_i = \eta_{i0}(1 + x\xi_0/b_i)$ is a solution to Eq. 19 as $x \to \infty$. As pointed out by deGennes [18], the linear dependence of the order parameter on $x$ appears to give an unphysical result as $x \to \infty$, however nonlinear terms neglected in Eq. 19 will introduce a negative curvature to the order parameter so that it will achieve its bulk value for $\xi_0 x > \xi(T)$. We are interested in the region $\xi_0 x \approx \xi_0 << \xi(T)$ where this curvature is negligible. To find the coefficient $b_i/\xi_0$ we use the variational approach of Svidzinskii [20] as it is presented by Samokhin [10] and by Barash *et.al.* [12]. Substituting $\eta = C(x + q(x))$ (then $b/\xi_0 = \lim_{x\to\infty} q(x)$) into the Eq. 19 gives

$$q(x) = \frac{E(x)}{2} + \int_0^\infty \tilde{K}(x,x') q(x') dx' \qquad (20)$$

with $E(x) = 2\int_0^\infty x' \tilde{K}(x,x') dx' - 2x$. The above equation can be found by minimizing the functional

$$\Psi[q] = \frac{\int_0^\infty dx\, q(x)[q(x) - \int_0^\infty dx'\, \tilde{K}(x,x') q(x')]}{[\int_0^\infty dx\, q(x) E(x)]^2}. \qquad (21)$$

The minimum value of $\Psi[q]$ is given by

$$\Psi_{min} = \frac{1}{2\int_0^\infty dx\, q(x) E(x)}. \qquad (22)$$

The coefficient $b$ can be expressed in terms of $\Psi_{min}$ as

$$\frac{b}{\xi_0} = \frac{\frac{1}{2}\int_0^\infty dx\, x E(x) + \frac{1}{4\Psi_{min}}}{\frac{1}{2}\int_0^\infty dx\, E(x) - \int_0^\infty dx'\, x' [\int_0^\infty dx\, \tilde{K}(x',x) - 1]}. \qquad (23)$$

Using a constant for $q(x)$ gives the following result

$$\frac{b_i}{\xi_0} = \frac{(7\zeta(3))^{-1}}{\int_0^1 s^2 F(s) ds + \frac{1}{2\pi} tr\left[\int_0^1 s \hat{\Delta}_i^\dagger(s) \times \int_0^1 s^2 \hat{\Delta}_i(-s) - \int_0^1 s^2 \hat{\Delta}_i^\dagger(s) \times \int_0^1 s \hat{\Delta}_i(-s)\right]} \qquad (24)$$

$$\times \left\{ \frac{\pi^4}{24}\left[\int_0^1 s^3 F(s) ds + \pi^{-1} tr \int_0^1 s^2 \hat{\Delta}_i^\dagger(s) \times \int_0^1 s^2 \hat{\Delta}_i(-s)\right] \right.$$

$$\left. + \frac{(7\zeta(3))^2}{2\pi^2} \cdot \frac{\left[\int_0^1 s^2 F(s) ds + \pi^{-1} tr \int_0^1 s \hat{\Delta}_i^\dagger(s) \times \int_0^1 s^2 \hat{\Delta}_i(-s)\right]^2}{\int_0^1 s F(s) ds - \pi^{-1} tr \int_0^1 s \hat{\Delta}_i^\dagger(s) \times \int_0^1 s \hat{\Delta}_i(-s)]} \right\}$$

where $\zeta(3) = \sum_n 1/(2n+1)^3$.

Values for $b_i/\xi_0$ are given in Table 1 for various functions $\Delta(\mathbf{k})$ corresponding to irreducible representations of $D_{6h}$ for a surface normal along the hexagonal **a** direction. For the case of $p$-wave pairing an exact solution can be compared to the variational solution. The $p$-wave order parameter transforms as a vector under spacial rotations. For the order parameter component transverse to the surface normal the variational solution gives $b_t/\xi_0 = 0.53$ which compares favorably to the exact result $b_t/\xi_0 = 0.54$ [13]. For the longitudinal component the order parameter obeys $\eta_l(0) = 0$ irrespective of the form of the boundary [13]. In this case the variational approach gives $b_l/\xi_0 = 0.11$ which is non-zero. This non-zero value arises because $\eta = C(b/\xi_0 + x)$ is an asymptotic solution to Eq. 19 and the exact boundary condition is valid only on the surface (see Fig. 1). The variational result for the constant order parameter ($b/\xi_0 = \infty$) is exact.

Note that in general $\xi(0) = \sqrt{\kappa/\alpha_0 T_c} \neq \xi_0$ (for isotropic superconductors $\xi(0) \approx 0.2\xi_0$ [18]). However $\xi(0) \approx \xi_0$ and the values of $b/\xi(0)$ ($\approx 0.6$) give rise to $H_{c_3} \approx H_{c_2}$ to less than a tenth of a percent within the temperature range shown in Fig. 2. This indicates that diffusive scattering effectively completely suppresses surface superconductivity for all but isotropic order parameters. Since the electronic wavelength is typically on the order of atomic length scales a surface will usually be diffusive. An interesting implication is that surface superconductivity is expected only to occur on surfaces with normals along high symmetry directions, where atomically clean surfaces can be cleaved. These results are consistent with the observation by Keller *et. al.* that cutting the crystals destroyed surface superconductivity in UPt$_3$ [8]. It has been proposed that turning a cylindrical superconductor in a magnetic field orthogonal to the axis of symmetry and measuring the surface superconductivity can determine the symmetry of the order parameter [10]. Our results indicate that such an experiment is not feasible because the surface of the sample will be diffusive.

We acknowledge the support of the Natural Sciences and Engineering Research Council of Canada. We also thank N. Keller, J.L. Tholence, A. Huxley, and J. Flouquet for making their work available to us prior to publication.

TABLES

| $\Gamma$ | $\hat{\Delta}(\mathbf{k})$ | $\frac{b}{\xi_0}$ specular | $\frac{b}{\xi_0}$ diffusive |
|---|---|---|---|
| $A_{1g}$ | $i\sigma_y$ | $\infty$ | $\infty$ |
| | $i\sigma_y(k_x^2 + k_y^2)$ | $\infty$ | 12 |
| | $i\sigma_y k_z^2$ | $\infty$ | 1.4 |
| $A_{2g}$ | $i\sigma_y(k_x^3 - 3k_x k_y^2)(k_y^3 - 3k_x^2 k_y)$ | 0 | 0.72 |
| $B_{1g}$ | $i\sigma_y k_z(k_y^3 - 3k_y k_x^2)$ | $\infty$ | 0.66 |
| $B_{2g}$ | $i\sigma_y k_z(k_x^3 - 3k_x k_y^2)$ | 0 | 0.63 |
| $E_{1g}$ | $i\sigma_y k_z k_x$ | 0 | 0.64 |
| | $i\sigma_y k_z k_y$ | $\infty$ | 0.46 |
| $E_{2g}$ | $i\sigma_y 2k_x k_y$ | 0 | 0.64 |
| | $i\sigma_y(k_x^2 - k_y^2)$ | $\infty$ | 1.3 |
| $A_{1u}$ | $i\sigma_y \boldsymbol{\sigma} \cdot \hat{z} k_z$ | $\infty$ | 0.53 |
| $A_{2u}$ | $i\sigma_y \boldsymbol{\sigma} \cdot \hat{z} k_z(k_x^3 - 3k_x k_y^2)(k_y^3 - 3k_x^2 k_y)$ | 0 | 0.68 |
| $B_{1u}$ | $i\sigma_y \boldsymbol{\sigma} \cdot \hat{z}(k_y^3 - 3k_y k_x^2)$ | $\infty$ | 0.72 |
| $B_{2u}$ | $i\sigma_y \boldsymbol{\sigma} \cdot \hat{z}(k_x^3 - 3k_x k_y^2)$ | 0 | 0.68 |
| $E_{1u}$ | $i\sigma_y \boldsymbol{\sigma} \cdot \hat{z} k_x$ | 0 | 0.11 |
| | $i\sigma_y \boldsymbol{\sigma} \cdot \hat{z} k_y$ | $\infty$ | 0.53 |
| $E_{2u}$ | $i\sigma_y \boldsymbol{\sigma} \cdot \hat{z} 2k_x k_y k_z$ | 0 | 0.58 |
| | $i\sigma_y \boldsymbol{\sigma} \cdot \hat{z}(k_x^2 - k_y^2)k_z$ | $\infty$ | 0.67 |

TABLE I. Boundary conditions for gap matrices transforming as selected basis functions of the point group $D_{6h}$. The surface normal is along the hexagonal **a** direction. The specular boundary condition is found by applying $P_\mathbf{x}\hat{\Delta}(\mathbf{k}) = \hat{\Delta}(\mathbf{k} - 2\mathbf{k} \cdot \hat{x}\hat{x})$. If $P_\mathbf{x}\hat{\Delta}(\mathbf{k}) = \hat{\Delta}(\mathbf{k})$ then $b = \infty$ and if $P_\mathbf{x}\hat{\Delta}(\mathbf{k}) = -\hat{\Delta}(\mathbf{k})$ then $b = 0$. The $\sigma_i$ are the Pauli matrices.

FIGURES

FIG. 1. Schematic representation of the extrapolation length $b$. (a) Microscopic depiction of the spacial variation of the superconducting order parameter near a superconductor to insulator boundary. (b) Macroscopic representation of (a).

FIG. 2. $H_{c_3}/H_{c_2}$ as a function of reduced temperature for various values of $b/\xi(0)$.